\documentclass[seceq]{ptptex}

\title{AdS-CFT and the RHIC fireball
}

\author{Horatiu \textsc{Nastase}%
}

\inst{Global Edge Institute, Tokyo Institute of Technology, Tokyo 152-8550, Japan
}

\abst{In this talk I will review my work on the description of high energy scattering in QCD, in particular the 
fireball observed at RHIC, as well as predictions for the LHC. The aim is to see how much we can learn about 
actual QCD (nonsupersymmetric, $N_c=3$), without knowing the details of the gravity dual of QCD. Experimental 
predictions are consistent with data, and important consequences are obtained for the LHC, in particular 
for the $pp$ collisions. The RHIC and LHC correspond to the regime of Froissart bound saturation,
in the Heisenberg model. Asymptotically, the RHIC fireball is mapped to a dual black hole in the IR 
of the dual. A simple (and unique)
scalar field theory model for the RHIC fireball indeed exhibits the properties of the 
dual black hole: a thermal horizon and aparent information loss. 
}

\newcommand{\be}{\begin{equation}}
\newcommand{\ee}{\end{equation}}
\newcommand{\bea}{\begin{eqnarray}}
\newcommand{\eea}{\end{eqnarray}}

\begin{document}

\maketitle

\section{Introduction}

It was hoped since the begining of AdS-CFT with Maldacena's paper \cite{malda}, that eventually we should be able to 
use the AdS-CFT correspondence to make real predictions about QCD, that can be experimentally tested, and 
in so doing have indirect experimental tests of string theory. Indirect since we would not be testing only 
string theory itself, but also the validity of the AdS-CFT correspondence that maps string theory to 
gauge theories. Of course, the major obstacle to that program is the lack of a gravity dual of QCD, even 
when the number of colours $N_c\rightarrow \infty$.

In this talk I will review work that I have done, trying to see if we can nevertheless extract some 
experimental QCD predictions for high energy colliders, with the limited AdS-CFT information that we have, 
i.e. without knowing the precise details of the gravity dual of QCD (but assuming that there is one).
Since fixed angle high energy scattering is the domain of perturbative QCD, which is well understood (not 
to mention hard to obtain with AdS-CFT since the gravity side would be nonperturbative), we have to look at 
small angle high energy scattering (high center of mass energy $s\rightarrow\infty$, but low momentum 
transfer $t$ in the $2\rightarrow 2$ scattering picture). This regime is known to be nonperturbative. 
We will find predictions for $\sigma_{tot,QCD}(s)$.

We will see that in this regime, asymptotically in the $s\rightarrow\infty$ limit, we create black holes in the 
gravity dual,
asymptotically living in the IR of the metric. 
In QCD, I will argue that that object corresponds to the fireball 
observed at RHIC, which thus is mapped one to one to an approximately 4 dimensional black hole, as first 
argued in.\cite{nastase} I will then describe the field theory implications of this idea.


Over the last few years there have been a large number of papers dealing with large $N_c$ 
${\cal N}=4$ SYM at finite temperature, thus in thermal equilibrium, as a model for the RHIC fireball, in 
which case a lot of information can be obtained. However, in that case one has to justify why do the results apply 
when the starting point is quite different from QCD. I will not discuss these developments.

\section{High energy QCD scattering from dual black hole production}

I will begin by analyzing small angle high energy scattering, first in flat space, and then in gravity duals.
In $2\rightarrow 2$ flat space scattering 
in a gravitational theory (like string theory) at 
fixed $t$, $s\rightarrow \infty$, 't Hooft showed long ago \cite{thooft} 
that if we are below the Planck scale, $\sqrt{s}\leq
M_{Pl}$, we can describe the quantum scattering in a simple quasiclassical way. One particle creates a 
gravitational shockwave of the Aichelburg-Sexl (AS) kind, 
\bea
ds^2 &=& 2dx^+ dx^- +(dx^+)^2 \Phi (x^i) \delta (x^+) +d\vec{x}^2\nonumber\\
&&\Delta_{D-2} \Phi (x^i) = -16 \pi G p \delta ^{D-2}(x^i)\label{lapl}
\eea
while the second moves on a null geodesic in this geometry. The resulting S matrix gives the correct 
differential cross section. In a curved space, like a gravity dual, we need to put the metric and 
Laplacean in (\ref{lapl}) in the curved background \cite{nastase5}. 

't Hooft also noted that if we are above the Planck scale, $\sqrt{s}>M_{Pl}$, then nonlinearity will imply that 
both particles will create gravitational shockwaves, in which case the scattering is described in a completely 
classical manner. The collision of two gravitational shockwaves in general 
relativity is a highly nonlinear process, and we expect to form a black hole. However, the future of such a 
collision is prohibitively difficult to calculate, a very complicated calculation of D'Eath and Payne \cite{dp}
being only able to find the metric as a perturbation near the collision point.
Nevertheless,  Penrose has shown that 
for an AS collision at zero impact parameter, $b=0$, 
a black hole does form. He showed that a marginally trapped surface 
forms at the point of collision which means, by a general relativity theorem, that a horizon will form outside
of the trapped surface, in the future of the collision. 

In 2002, Eardley and Giddings \cite{eg} extended Penrose's formalism to flat $D=4$ space with $ b>0$, finding the maximum $b$ 
for which a trapped surface forms. Then 
\be
\sigma_{BH\; prod.} \geq \pi b_{max}^2 (s)\label{bmax}
\ee
Yoshino and Nambu \cite{yn} found a numerical method to extend this formalism to flat $D>4$ space. In \cite{kn3}, 
we found an approximate method to calculate $b_{max}(s)$ in (\ref{bmax}), 
that can be extended to curved $D>4$ space, arbitrary shockwave profile $\Phi$
and $b>0$, thus being able to apply it to gravity duals. We also calculated string corrections to the method, by 
using a string corrected shockwave profile $\Phi$ found by Amati and Klimcik \cite{ak} 
from the same 't Hooft-like set-up.
The result was that string corrections are exponentially suppressed at $s\rightarrow \infty$, as 
$e^{-s/s_0}\rightarrow 0$. This was in line with general arguments made by 't Hooft in the original paper \cite{thooft},
but we wanted to see explicitly that in the $s\rightarrow \infty $ limit (with $t$ fixed), the classical 
gravity calculation gives the right result. 
This formalism was then applied to gravity duals in \cite{kn,nastase2}. 

Of course, we do not know the gravity dual of QCD. Assuming that there is one, we 
know that it looks like AdS space in the UV of the metric (at large $r$), since QCD is approximately 
conformal at large energies. Note that now large energies and UV refers to fixed angle scattering, i.e. that 
we scale up all the momenta by the same amount. We also know that the gravity dual terminates smoothly in the 
IR (without singularities). The simplest model for a gravity dual that we can take is 
(hiding our ignorance about the way in which the space terminates) $AdS_5$ space with a
cut-off, at a position related to the scale of the lightest glueball state in QCD, $\bar{r}_{min}=R^2\Lambda$,
\be
ds^2= \frac{\bar{r}^2}{R^2} d\vec{x}^2 + \frac{R^2}{\bar{r}^2}d\bar{r}^2 + 
R^2 ds_X^2= e^{-2y/R}d\vec{x}^2 + dy^2 + R^2 ds_X^2
\ee

This is the model considered in 2001 by Polchinski and Strassler \cite{ps}, in order to define the gravity dual 
of high energy QCD scattering. The amplitude for scattering in QCD is found by convoluting the string 
scattering amplitude in the dual with wavefunctions for the external states (integrated over the extra 
dimensions), while rescaling the momenta by the warp factor, $\tilde{p}_{(AdS)}=\frac{R}{\bar{r}}p_{(QCD)}$.

Then, energy scales for scattering in the gravity dual correspond to energy scales in QCD. In particular, 
the Planck scale $M_P$ in the dual corresponds to $\hat{M}_P=N_c^{1/3}\Lambda_{QCD}$ in QCD, a scale above 
which we create black holes in the dual. The small angle high $s$, fixed $t$, scattering is nonperturbative, 
but governed by IR (low energy) physics, since it is dominated by the emission of many low energy pions (the 
lowest mass particles). This fact manifests itself most readily in the maximum possible behaviour, the 
saturation of the Froissart unitarity bound for $\sigma_{tot, QCD}(s)$, which is governed by the mass of the pion
$m_{\pi}$ (lowest mass QCD excitation).

It is then not so surprising that we find (using the Polchinski-Strassler formalism and 
our black hole production formalism) that as $s\rightarrow \infty$ in QCD, in the dual we produce black holes 
that are closer and closer to the IR of the metric, i.e. to the cut-off $\bar{r}_{min}$. The limiting 
behaviour that we find is of black holes effectively on the IR cut-off, and then \cite{kn}
\be
\sigma_{tot,QCD}(s)\sim \bar{K}\frac{\pi}{M_1^2}\ln^2 \frac{s}{s_0}
\ee
where $\bar{K}$ is a constant depending on the model for cut-off in the IR, and $M_1$ is the lightest 
excitation: KK reducing gravity onto the IR cut-off ("IR brane"), the lowest
mass is $M_1\simeq 3.83/R_{AdS}$, corresponding to the lightest QCD excitation.

There is an effective scattering region in which most of the integration over the extra dimensions in the PS 
formula is located. When $s$ is comparatively small ($\hat{M}_P<\sqrt{s}<\hat{E}_R=N_c^2 \Lambda_{QCD}$), 
the dual black holes are smaller than this region, therefore they have quantum fluctuations inside it (due to 
the PS integration). When $\hat{E}_R<\sqrt{s}<\hat{E}_F=?$, the black holes are bigger than the 
scattering region, thus are almost classical. Finally, above the unknown scale $\hat{E}_F$ (corresponding to Froissart 
bound saturation), in the dual the black holes are effectively "stuck" on the IR brane (cut-off). The value of 
$\hat{E}_F$ can only be calculated once we know the details of the IR of the QCD dual metric.
Since string corrections in the dual are exponentially small 
as $s\rightarrow\infty$, we can apply the above formalism 
to finite $N_c$ and $g^2N_c$, thus to QCD. In QCD we have the additional 
complication that the lightest state is the pion, which is not a glueball. A simple model for the dual of the pion 
in this set-up is the fluctuation of the IR cut-off. In other words, make the cut-off a real IR brane, in which 
case it also gives the dual of the Froissart bound, as gravity did. 

The QCD energy scales relevant for $\sigma_{QCD,tot}(s)$ are then: The first is the glueball scale, approximately 
equal to the gauge theory "Planck scale", $\Lambda_{QCD}\sim \hat{E}_S\sim \hat{M}_P=\Lambda_{QCD}
N_C^{1/4}\sim 1-2 GeV$, where the experimental value is the estimated value for the lightest glueball mass.
The second is the scale $\hat{E}_R=N_c^2\Lambda_{QCD} \sim 10 GeV$ and the third is the Froissart saturation 
scale, $\hat{E}_F$. From experimental data $\hat{E}_F<1 TeV$, since $\sigma_{tot}^{QCD}(s)$ reaches 
$\pi/m_{\pi}^2$ (the coefficient of $\ln^2 s$) before $1TeV$. We then find from black hole production in 
the dual the prediction that:\cite{nastase2}

{\bf Between $\Lambda_{QCD}$ and $\hat{E}_R$, $\sigma_{tot}\sim s^{1/7}$; between 
$\hat{E}_R$ and $\hat{E}_F$ we have 

$\sigma_{tot}\sim s^{1/11}\simeq s^{.0909}$; above $\hat{E}_F$,
the saturation of the Froissart bound.}

Experimentally, we have the "soft Pomeron" behaviour, onsetting at $9 GeV$
\be
\sigma_{tot,AB}=X_{AB} s^{\epsilon}; \;\;\; \epsilon = 0.0933\pm 0.0024;\;\;\;\;
X_{AB}\sim 10-35 mb
\ee
with a $\chi^2/d.o.f=1$, thus in complete agreement with my prediction. 
However, later it was argued that a slightly better fit is 
\be
\sigma_{tot}=Z_{AB}+B\ln^2 \frac{s}{s_0};\;\;\;\;
Z_{AB}\sim 18-65 mb;\;\;\; B=0.31 mb\ll 60 mb
\ee
valid down to $\sqrt{s}=5GeV$, with $\chi^2/d.o.f.=.971$. But I would argue that this is just a coincidence,
since we get this by expanding $s^\epsilon$ and dropping the linear term, which explains much better the 
unnatural relation $\pi/m_{\pi}^2= 60mb\gg B$, by the fact that $B/Z_{AB}\sim \epsilon^2\sim 0.01$.
Also note that if we take the onset of the "soft Pomeron" behaviour as the experimental value of $\Lambda$, 
$\Lambda_{QCD, exp.}=9 GeV/N_c^2=1GeV$, then we get $M_{P,exp.}=N_c^{1/4}\Lambda_{QCD,exp.}=1.3GeV$.  

Moreover, there is some evidence of thermality of the high energy scattering total cross section in the 
diffractive vector meson (VM) photoproduction cross section, though the error bars are still large.
One can argue that if we extrapolate the VM photoproduction cross section down to the Planck scale,
$\sqrt{s}=M_P$, we expect the ratios there for cross sections of different vector mesons 
to satisfy\cite{fn}
\be
\sigma_{VM}|_{\sqrt{s}\sim M_P}\propto |<0|j_{\mu}|V^a>|^2\exp\left(-\frac{4M_V}{T_0}\right)
\ee
This is experimentally correct within error error bars, and from the data one finds $
T_0\simeq 1.3 GeV\simeq M_{P,exp.}$, the "experimental gauge theory Planck scale" above.
More precisely, we predict for the vector meson V
\be
\sigma_V\propto |<0|j_{\mu}|V^a>|^2\left[\exp\left(-\frac{4M_V}{T_0}\right)\right]\left(\frac{s}{M_P^2}
\right)^{\epsilon_V}\label{scalin}
\ee
where $\epsilon_V$ depends on the vector meson. For the heavy vector mesons, $\epsilon_V$ is much different 
than the "soft Pomeron" value valid for light mesons, on general grounds. In particular, we predict for the 
extrapolation of the (\ref{scalin}) power law down to $1 GeV$ 
\be
\frac{\sigma_{J/\Psi}}{\sigma_{\Upsilon}}|_{\sqrt{s}\simeq M_P}\sim 10^9
\ee
This is consistent with the current data, but the current error bars are huge. However, due to the large ratio, 
this is potentially a very good test of this picture, and hopefully the LHC will be able to test for it.

So what else can we predict for the LHC? First, and most importantly, I would hope that their sensitivity for the 
total cross section will increase, so they can distinguish between various curves, as my prediction for
$\sigma_{tot,QCD}(s)$ is very specific. We have seen that there is only a weak dependence in this $s\rightarrow\infty$
small angle scattering on the particles A and B that are being scattered (for $\sigma_{QCD, AB}^{tot}(s)$ only the 
coefficient in front depends on them). In particular, the physics is the same: thermal creation of dual black 
holes. We will develop this picture further, and see that the black holes created in the IR of the dual are 
related to the RHIC fireballs, of average temperature $T\sim 175 MeV$. But since the same physics should apply 
according to the picture discussed here, in the {\em $pp$ collisions at the LHC we expect the same thermal 
effects}, specifically that 
\be
T_{pp,LHC}\simeq T_{Au+Au, RHIC}\simeq T_{phase \;\;transition}\simeq 175 MeV\label{tempe}
\ee

\section{The 1952 Heisenberg model 
for Froissart saturation from AdS-CFT}

In 1952, even before the advent of QCD, let alone the Froissart bound, Heisenberg \cite{heis} proposed a simple model 
that gives the saturation of the Froissart bound. In \cite{kn2}, we found that the same model is obtained from 
the AdS-CFT description from the previous section.

The Heisenberg model starts with the observation that at high energy, both the colliding hadrons and the 
(effective) pion field surrounding them Lorentz contracts. Heisenberg argues that as $s\rightarrow\infty$, 
the hadrons become irrelevant, except as sources for the Lorentz contracted pion field, thus in the 
asymptotic limit, we can describe the process as just a scattering of pion field shockwaves. 
The energy loss should be proportional to the total energy of the collision and the overlap of the 
pion field wavefunctions, thus he obtains
${\cal E}\sim  e^{-bm_{\pi}}\sqrt{s}$.
The minimum energy loss corresponds to ${\cal E}=<E_0>$, the average emitted energy per pion, in which case we obtain 
the maximum impact parameter that gives hadron interaction, $b_{max}$. Then 
\be
\sigma_{tot}=\pi b_{max}^2(s)=
\frac{\pi}{m_{\pi}^2}ln^2\frac{\sqrt{s}}{<E_0>}
\ee
i.e., if $<E_0>$ is constant, the saturation of the Froissart unitarity bound. However, for a usual $\lambda\phi^n$
interaction (or for free fields), $<E_0>\propto \sqrt{s}$ and we don't find saturation. But Heisenberg argues that 
we need to take a purely derivative interaction for the pion, for which he takes as a prototype the DBI action
\be
S_{DBI}= l^{-4}\int d^4x \sqrt{1+l^4[(\partial_{\mu} \phi)^2 +m^2\phi^2]}\label{dbi}
\ee
and finds that for it indeed $<E_0>\simeq m_{\pi}\frac{\ln \gamma}{1-1/\gamma}\simeq$ constant. But this is nothing 
other than the action of radion (fluctuation of the IR brane), 
${\cal L}=l_s^{-4}\sqrt{\det (\partial_a X^{\mu}\partial_b X^{\nu}g_{\mu\nu})}$,
in the simple model for the pion that we took in the previous section, with $m=0$.
Thus the radion action is needed in order to get the saturation of the Froissart bound. We argued that in the 
physical case when the pion is lightest, the fluctuations of the IR brane give the same physics as gravity, and
the Froissart bound is saturated. 

However, in the purely gravitational case (dual to pure Yang-Mills theory without quarks), we had the exact same 
picture as Heisenberg, replacing the collision of pion field shockwaves with the collision of gravitational 
field shockwaves, effectively living (in the $s\rightarrow \infty$ limit) on the 4 dimensional IR brane. 
The "wavefunction" of the graviton, the shockwave profile $\Phi$, did not need to be postulated like 
Heisenberg did, but was derived, as was the maximum impact parameter $b_{max}(s)$
\be
\Phi=R_s\sqrt{\frac{2\pi R}{r}}C_1e^{-M_1r};\;\;\; M_1=\frac{j_{1,1}}{R}\sim \frac{3.83}{R}
\ee
so in a sense gravity is easier than scalar field theory. The mass term inside the square root in 
(\ref{dbi}) is a nonlinear generalization of radion stabilization.

We can conclude that the DBI action is a good model action for the small angle $s\rightarrow\infty$ scattering. 
But in the gravitational description dual to pure Yang-Mills, for $\sqrt{s}>\hat{E}_F$ an 
almost classical black hole 
is created and decays, approximately on the 4d IR brane. Therefore in QCD we also expect a 
pion field thermal solution to be created and to decay, and I will argue that this is nothing but the RHIC 
fireball!\cite{nastase,nastase4} It also means that the DBI model action must admit such a solution.

\section{The RHIC fireball as a dual IR black hole (approximately 4d)}

At RHIC, we have $Au+Au$ (A=197) collisions at about $100 TeV$ per nucleon, or about $40 TeV$ total energy, 
surely more than $E_F$, which we argued is $<1 TeV$, therefore the experiment should be in the regime of Froissart 
bound saturation. Therefore we would expect to produce black holes on the IR brane in the dual, and 
correspondingly to produce a thermal pion field ("fireball") solution in QCD. But we do observe a fireball, 
made up of strongly coupled quark-gluon plasma, or sQGP.

It decays thermally, at an average temperature that is constant (independent of the mass of the fireball), 
and approximately equal to the (lattice value for the) phase transition temperature, 175 MeV. But the 
dual black hole on the IR branes also has an approximately constant value \cite{nastase,nastase4}. 
This picture of the black hole 
on the IR brane has been developped further in \cite{amw} under the name "plasma ball".

Note that the picture described here is for high energy scattering in actual QCD (at zero "outside" temperature, 
the fireball is an finite temperature object unstable against decay),
not for "${\cal N}=4$ SYM at finite temperature", as it has become popular in the last few years. We were able to say
something about QCD, since string $\alpha'$ and $g_s$ corrections in the dual are exponentially small for 
this process, and that translates to small $1/N_c$ and $1/(g^2N_c)$ corrections for QCD.
We are looking at $s\rightarrow\infty$ and $t$ fixed, or rather at the "soft" processes dominating the cross 
section, with many emitted particles of small energy. Since $\sigma_{tot}(s)\sim Im A(s,t=0)$,
the two pictures (fixed and small t, or processes dominating the total cross section) are equivalent.

We now emphasize several unusual properties of the dual gravity description:

\begin{itemize}
\item Dual gravity is effectively 4 dimensional, as one KK reduces it to the IR brane, where the scatterring 
effectively takes place. Gravity is dual to glueball fields.
\item The reduced gravity is massive. If we reduce on an UV brane in the RS model, gravity is massless, but 
by reducing onto the IR brane, we get a mass (lowest KK mode) of $M_1=j_{1,1}/R\simeq 3.83/R$.
\item Then the dual reduced gravity acts on a short distance only, by $V(r)\sim e^{-M_1r}/r$.
\item The dual black hole has a horizon only, but no singularity. This was argued in. \cite{amw} 
The AdS space cuts off at $\bar{r}_{min}$, outside the black hole, whereas the would-be singularity of the 
(infinitely extended) black brane would be at $\bar{r}=0$. 
\item The dual of the pion field is the position of the IR brane, and it has a similar physics as gravity 
in the physical case of $m_{\pi}<m_{glue}$.
\item The pion action needs a solution with thermal solution, analog of the black hole
\end{itemize}

Now we try to compute the temperature of the fireball from the temperature of the dual black hole.
Working first in the pure gravity case, from $dM=TdS=TM_{P,4}^2 d(Area)/4=\pi TM_{P,4}^2dl_H^2/4$, we 
get that if we have the scaling relation
\be
dl_H^2=\frac{dM}{aM_1} \label{scaling}
\ee
then we obtain for the dual black hole 
\be
T_{dual\; BH}=a\frac{4M_1}{\pi}
\ee
Since $M_1$ is the lightest physical excitation, it is replaced by $<m_{\pi}>$ in QCD, thus if $a=1$
(we have not computed $a$, and we cannot until we have the 
black hole on the IR brane solution), we would get a value very close to the experimental one,
\be
T_{fireball}=4\frac{<m_{\pi}>}{\pi}=175.76 MeV
\ee
But why do we have the scaling relation (\ref{scaling})? There are several reasons.

1) A scaling relation, obtained by using Hawking's asymptotic flat space relation $T=\kappa/(2\pi)$. We find 
$\kappa\simeq M_1$, thus (\ref{scaling}) is obtained.

2) A dimensional argument: gravity acts at short distance now, $V\sim e^{-M_1 r}/r$, thus we expect $M_1$ to 
dominate low energy physics.

3) An argument related to the physical properties of the dual black hole. It has a finite extent and a 
horizon with a pancake-like shape, stretched along the IR brane. The horizon is nothing special, so no 
strong curvatures are felt when crossing it. But inside it, one finds a region of strong curvature, of a 
similar pancake shape, where gravity is fully five dimensional. Thus to calculate the 4 dimensional 
size of the black hole, we can use the region of strong 5d gravity (as viewed from the 
outside) as a guide. We can use 
5d Newtonian gravity, and approximate the boundary of this region as the region where the Newtonian potential 
becomes of order one. Thus
\be
h_{00}\sim \frac{M}{M_{P,5}^3r^2}\propto \frac{M}{M_{P,4}^2M_1r^2}\sim 1\Rightarrow 
M_{P,4}^2r_H^2= \frac{M}{aM_1}
\ee

4) We also have a purely pion field argument that will be presented in the next section,
for the corresponding scaling relation in field theory, 
\be
\hat{M}_P^2r_H^2\propto \frac{M}{m_{\pi}}\label{scalingf}
\ee

Towards the identification of the fireball with the black hole in the IR of the dual, we can offer some 
qualitative arguments as well. Firstly, as is well known, the ratio of the shear viscosity over entropy density, 
$\eta/s$, is argued to be bounded from below by $1/(4\pi)\simeq 0.1$, and the experimental value at 
RHIC (as you have heard from other talks) is very close to the minimum value (within experimental and 
theoretical errors). Moreover, it was found \cite{bl} that in theories that have a gravity dual (thus 
in QCD itself, not just in ${\cal N}=4$ SYM) , dual black holes saturate the minimum bound. Therefore, 
if the above picture is correct, the fireball should have $\eta/s=1/(4\pi)$, which is true within current errors.

Secondly, the phenomenon of "jet quenching", that jets (hard scattering) are strongly suppressed, can be understood 
as the dual black hole information loss: information goes in (transverse momentum of hard particles), but 
only thermal radiation comes out. Of course, jet quenching is not perfect, and in fact from ${\cal N}=4$ 
SYM at finite temperature people have been able to calculate the jet quenching parameter in various models, 
but the information loss in this model is not perfect as well. Indeed, the black hole is only approximately 
classical, and it quantum fluctuates (is integrated) over the extra dimensions, allowing an "escape route"
for the incoming hard particles. 

Thirdly, the temperature of the fireball equals the temperature of the phase transition 
in our model. Indeed, it was found in \cite{amw} that one needs to equate the temperature of the inside of the 
black hole, corresponding to the deconfined phase in gauge theory, with the temperature of the 
outside (confined phase), thus being at the temperature boundary of the two phases.

\section{A simple model: BIons with thermal horizons, "pionless holes"}

I argued that we need to find a thermal pion field solution as a solution of the good model action for the 
pion, the DBI action. Indeed, the DBI action $S=\int \sqrt{1+(\partial \phi)^2} $ describes the fluctuations
of the IR brane dual to the pion, and was needed by Heisenberg to obtain saturation of the Froissart bound, 
so it should describe fireballs.
But indeed the DBI action does admit solutions of the type we want, specifically the "catenoid" solutions. 
They are analogs of the BIon solution for the original 
BI action for the electric potential $\phi$. The catenoid solution is
\be
\phi'(r)=\frac{\bar{C}}{\sqrt{r^4-\bar{C}^2}}\Rightarrow \phi(r)=\bar{C}\int_r^{\infty} \frac
{dx}{\sqrt{x^4-\bar{C}^2}}\rightarrow \frac{\bar{C}}{r}\label{catenoid}
\ee

Note that the catenoid has an apparently singular "horizon"-like structure at
$r=r_0=\sqrt{\bar{C}}$, i.e. a place where 
$\phi '(r)\rightarrow \infty$, but where $\phi(r)$ remains finite. At this point this is just a 
formal observation, but we will see that this horizon is thermal and absorbs information.\cite{nastase3}
Also note that since $\bar{C}$ is a quantized scalar charge, as we can see from the behaviour 
at infinity of (\ref{catenoid}), we have ($\Lambda$ is the DBI scale)
\be
\hat{M}_P^2r_0^2\sim \Lambda^2r_0^2=\bar{C}=\frac{M}{m_{\pi}}\label{scalcat}
\ee
which implies the field theory scaling relation (\ref{scalingf}), as advertised in the previous section.
Thus we have a purely field theory argument for $T=4a m_{\pi}/\pi$.

However, before we analyze the properties of the horizon, we must notice that unlike for the black hole, 
one cannot continue the catenoid solution inside this horizon to a singularity. The only way we can 
continue the solution is to a different asymptotic region, interpreted as an anti-D brane. Since the 
scalar $\phi$ is dual to the position of a D-brane, $\phi'\rightarrow\infty$ and $\phi$ finite means that at the 
horizon the dual D-brane is perpendicular to the asymptotic region, thus can be glued onto another solution 
with a different asymptotic region, forming a $D-\bar{D}$ system joined by a thick string. But since 
we cannot create a whole new D-brane in the process of scattering two hadrons on the original D-brane, the only 
possibility is that we create a large metastable "bubble", like a baby Universe joined by a throat to the 
original one. 

So the catenoid, which is a perfectly allowed solution outside the horizon, can only be created by scattering 
two hadrons, in which case we create a large metastable bubble looking like the catenoid near the horizon.

We want now to understand how it is possible to have a thermal horizon in quantum field theory. It turns out 
that there is a simple model with the same properties, that in fact can be mapped to our scalar field theory model, 
the "dumb holes" found by Unruh.\cite{unruh} These are obtained in a  nonrelativistic irrotational 
($\nabla \times \vec{v}=0$) fluid flow moving at ultrasonic speeds (see also \cite{volovik} for other 
nonrelativistic models). There the surface of $v=c$  acts as the thermal horizon of a black hole. 

The parameters of the flow are the speed of the flow, $\vec{v}=\vec{\nabla}\Phi$, the density 
$\rho$ and the sound speed $c^2=dp/d\rho$. Then Unruh showed that the 
fluctuation equation for the velocity potential, $\phi=\delta\Phi$, 
describing the fluid is
\be
\partial_0\frac{\rho/c^2}{1-v^2/c^2}\partial_0\phi+\partial_i\rho(\frac{v^iv^j}{c^2}-
\delta^{ij})\partial_j\phi=0
\ee
It matches the fluctuation equation for a scalar field in the black hole background 
\be
ds^2=\frac{\rho}{c}[(c^2-v^2)d\tau^2-(\delta^{ij}+\frac{v^iv^j}{c^2-v^2})dx^idx^j]\label{effmetric}
\ee
One can then follow Hawking's calculation step-by-step in this effective metric, to 
obtain that the the temperature of the thermal horizon at v=c is
\be
T=\frac{k}{2\pi}=\frac{1}{4\pi}\frac{1}{\rho}\partial_r[\rho c(1-v^2/c^2)]|_{v=c}
\ee

The horizon $r=r_0$ of the catenoid solution of the DBI action is in fact exactly like Unruh's, 
even though that case was non-relativistic. We can map the scalar fluctuation equation in the 
catenoid background to the Unruh case by 
\be
c^2=1+(\nabla \phi)^2; \;\;\; \rho = \frac{1}{\sqrt{1+(\nabla \phi)^2}}
;\;\;\; \vec{v}=\vec{\nabla}\phi
\ee
Unfortunately if we do this we obtain 
\be
T_{horizon}=\frac{k}{2\pi}\sim \partial_r g_{tt}\rightarrow \frac{dv}{dr}|_{v=c}\rightarrow
\infty
\ee
so we need some sort of regularization (introducing higher order corrections, time dependence - since 
the catenoid is supposed to be metastable anyway-, or non-sphericity, since the physical collision of 
hadrons is not spherically symmetric, are possible solutions). A good sign in that respect is the 
fact that if we introduce a small pion mass $m_{\pi}$, we obtain that $T\propto m_{\pi}$ (times an 
infinite constant), as expected.

However, a different type of modification is also possible.\cite{nastase6} In Heisenberg's model, the hadrons are not 
important except as a source for the pion field. It is then possible that we can model the presence of 
the hadrons by just a fixed source term added to the DBI action, $\int d^4 x \phi g(r)$, where $g(r)$ 
is a nucleon ($\bar{N} N$) distribution. There is no information to fix $g(r)$, but 
let us take the toy model $g(r)=\alpha/r^2$. Then 
\be
\phi(r)=\int_r^{\infty} dx \frac
{\bar{C}+\alpha x}{\sqrt{x^4-(\bar{C}+\alpha x)^2}}
\ee
Then if we take $\bar{C}=-\alpha^2/4$, the horizon is at $r_0=\alpha/2$, and at the horizon we have
\be
\phi (r)\simeq\frac{\alpha}{2\sqrt{2}}\ln (r-r_0)\rightarrow \infty \;\;{\rm as}\;\;
r\rightarrow r_0
\ee
In this case we do obtain a finite temperature, $T=\sqrt{2}/(\pi \alpha)=2\sqrt{2}/(\pi r_0) $.
For this action and solution we also have $\bar{C}=M/m_{\pi}$ = quantized scalar charge, and we also obtain
the desired scaling relation (\ref{scalcat}). The above solution has been dubbed "pionless hole"
by similarity with black holes and dumb holes for light and sound, respectively.
But now we have a horizon with a finite temperature, and we also obtain that the time it takes a 
perturbation to reach a point $r$ near $r_0$ is
\be
t\sim \frac{\alpha}{2\sqrt{2}}\ln (r-r_0)\rightarrow \infty 
\ee
thus the horizon also traps information, like the horizon of a black hole. But how is it possible to have 
information trapped at the horizon in a relativistic field theory without gravity? The answer is that 
unlike for a black hole, only scalar field perturbations are trapped, other kinds of information can travel freely. 
For $\delta\phi$, the phase and group velocities become zero
at the horizon (like in an extreme medium), 
\be
c_{ph}=\frac{\omega}{k}\propto r-r_0\rightarrow 0;\;\;\;\;
c_{gr}=\frac{d\omega}{dk}\propto \sqrt{r-r_0}\rightarrow 0
\ee

I have also proven that the unique relativistic action for a single scalar field that admits a solution with 
a thermal horizon (with finite T) that traps information is the DBI action with a fixed source term. 
The source function g(r) is only restricted to decay faster than a free field (1/r) at
infinity, but is otherwise arbitrary.\cite{nastase6}

So we have now a situation where the information loss paradox appears in the context of scalar field theory 
(albeit a unique one), since information is trapped at the horizon, and only thermal radiation comes out of 
the pionless hole. But how does this match with the statement that Hawking's derivation is for gravity 
only? In fact, Hawking's derivation uses the kinematics of gravity only, i.e. the fact that there is 
a black hole metric, but not the dynamics of gravity (Einstein's equations)
-in particular, we do not need to know the metric inside the 
horizon. But here we do have an effective black hole metric, namely (\ref{effmetric}), just that it is 
not defined by Einstein's equations, but instead is defined by the DBI scalar field equation. 

The conclusion is that the information loss paradox has nothing to do with gravity, as it appears here for a 
scalar field. Since the scalar field theory is unitary, it follows that the information must be 
retrieved in some way. The way to resolve the paradox would be to find a formalism that can 
deal with the process of formation and decay of a thermal object from the collision of two (T=0) 
quantum objects.


\section{Difference from Witten's thermal AdS-CFT and implications}

To illustrate the point, we will emphasize the differences\cite{nastase4} 
between the picture described in this talk and
the usual treatment of thermal AdS-CFT by Witten.\cite{witten}

Both in the usual treatment of thermal AdS-CFT by Witten and in the picture in this talk, the gravity dual,  
which in both cases is approximately AdS, contains a black hole. For an AdS black hole one can calculate 
the dependence of the temperature of the black hole as a function of its mass, and find that
$T(M)\sim M^{-1/2}$ at low M and $\sim M^{1/4}$ at high M, forming two branches, touching at a minimum $T=T_{min}$.

The first difference is that in \cite{witten} the black hole is in thermal equilibrium with 
AdS space (corresponding to a temperature constant over space and time in the field theory). Equilibrium is 
achieved only at large M, and in fact requires scaling $M$ and other quantities as 
$M\rightarrow\infty$,  so one is forced onto the $T(M)\sim M^{1/4}$ branch. 

But now we want a non-equilibrium situation, where a black hole is created and then decays, like the Schwarzschild 
black hole in flat space. In fact, the lower branch $T(M)\sim M^{-1/2}$ is the correct scaling for the 5 dimensional
(flat space) Schwarzschild solution, so that is the correct branch to consider now.

The second difference is that we are not in $AdS_5$ anymore, but in cut-off $AdS_5$ (toy model for QCD). As a 
consequence, the second branch dissappears completely, and instead the first (flat space) branch 
goes over smoothly to $T(M)\simeq T_0=$ constant at $M\rightarrow\infty$, corresponding to the 
asymptotic case of the black hole on the IR brane. 

Thus in order to describe the collision at $\hat{E}_F<\sqrt{s}\rightarrow\infty$
of two zero temperature hadrons  or nuclei in QCD, forming the finite temperature 
fireball, and then decaying, I argue that we need this different type of dual black hole at $T=T_0$, living on the 
IR brane. And then the black hole paradox is already present at RHIC.

Indeed, the collision of T=0 quantum objects (the Au nuclei), followed by the formation of the high T 
 fireball, and its decay, is exactly the same process as the collision of two dual hadrons, followed 
by the formation of the black hole on the IR brane, and its decay. Both can be described as giving an 
information loss paradox. 
Therefore the lack of a formalism that can deal with all the stages of the fireball at once (collision, thermal 
state and decay), instead of the usual formalism (say, the Matsubara formalism) for a fixed temperature over space and 
time, far from being a "minor inconvenience", is the very definition of the information loss paradox.

The best test for this aparent information loss at RHIC would be to measure the area law. For a black hole, 
the entropy is $S_{BH}=M_P^2 Area/4$. By what we said in this talk, for the fireball we expect 
then 
\be
S_{fireball}= \frac{\hat{M}_P^2}{4} {\rm Area}  \label{arealaw}
\ee
where the QCD "Planck scale" is $\hat{M}_P\sim 1-2 GeV$. The entropy of the fireball can't be measured 
directly, but we can measure the entropy of the emitted pions (which is larger, and proportional to it), 
by 
\be
S_{\pi}\simeq \frac{2\pi^4}{45 \zeta(3)}N_{\pi} \simeq 3.6 N_{\pi}
\ee

However the real problem is that we can't experimentally measure the area of the fireball when it starts 
decaying. If one would be able to measure it, we would have in the area law in (\ref{arealaw}) a 
perfect test for aparent information loss at RHIC.

What new phenomena do we expect at the LHC then? I have said in section 2 that the Froissart saturation regime 
will be reached in the $pp$ collisions at the LHC (since $14 TeV>1 TeV>\hat{E}_F$), therefore the LHC should 
see dual black holes on the IR brane, corresponding to the scalar field (pion) fireball described in this talk.
I have argued that the fireball at RHIC matches the properties of this scalar field fireball, but the 
incoming nuclei contain a number of nucleons, and this modified the observed result to some degree. 

Instead, I expect that the LHC will observe a much cleaner (purer) version of the scalar field fireball described 
here. Therefore, I predict that the $pp$ collisions at the LHC will also observe a thermal object, with 
a similar temperature, as in (\ref{tempe}).




\end{document}